\documentclass{llncs}
\usepackage{subfig}
\usepackage{graphicx}
\usepackage{amssymb}
\usepackage{ctable}
\usepackage{multirow}
\usepackage{dsfont}
\usepackage{tabularx}
\usepackage{bbm}
\usepackage{booktabs}
\usepackage{siunitx}

\usepackage{hyperref}
\hypersetup{
    colorlinks=true,
    linkcolor=blue,
    filecolor=magenta,      
    urlcolor=cyan,
}

\newcommand{\mcc}[1]{\multicolumn{1}{c}{#1}} 

\begin{document}

\title{Cardiac MRI Segmentation with Strong Anatomical Guarantees}
\titlerunning{Deep learning cardiac segmentation}  
%
\author{Nathan Painchaud\inst{1} \and Youssef Skandarani\inst{1,2,4} \and Thierry Judge\inst{1} \and Olivier Bernard\inst{3} \and Alain Lalande \inst{2} \and Pierre-Marc Jodoin\inst{1}}

\authorrunning{Nathan Painchaud et al.} 

\tocauthor{Nathan Painchaud, Youssef Skandarani, Thierry Judge, Olivier Bernard, Alain Lalande, Pierre-Marc Jodoin}
 
\institute{Department of Computer Science, University of Sherbrooke, Canada$^1$\\
Universit\'{e} de Bourgogne Franche-Comt\'{e}, Dijon France$^2$ \\
Universit\'{e} de Lyon, Lyon, France$^3$ \\ 
CArdiac Simulation \& Imaging Software$^4$}

\maketitle              

\begin{abstract}
Recent publications have shown that the segmentation accuracy of modern-day convolutional neural networks (CNN) applied on cardiac MRI can reach the inter-expert variability, a great achievement in this area of research.  However, despite these successes, CNNs still produce anatomically inaccurate segmentations as they provide no guarantee on the anatomical plausibility of their outcome, even when using a shape prior.  In this paper, we propose a cardiac MRI segmentation method which always produces anatomically plausible results.  At the core of the method is an adversarial variational autoencoder (aVAE) whose latent space encodes a smooth manifold on which lies a large spectrum of valid cardiac shapes.  This aVAE is used to automatically warp anatomically inaccurate cardiac shapes towards a close but correct shape.  Our method can accommodate any cardiac segmentation method and convert its anatomically implausible results to plausible ones without affecting its overall geometric and clinical metrics.  With our method, CNNs can now produce results that are both within the inter-expert variability and always anatomically plausible.
\end{abstract}

\section{Introduction}
Magnetic Resonance Imaging (MRI) is a non-invasive imaging technique of choice to evaluate the heart. The cardiac function is typically evaluated from a series of kinetic images (cine-MRI) acquired in short-axis orientation~\cite{Salerno17}.  In clinical practice, cardiac parameters are usually estimated from the knowledge of the endocardial and epicardial borders of the left ventricle (defined as the cavity (LV) and the myocardium (MYO)) and the endocardial border of the right ventricle (RV) in end-diastolic (ED) and end-systolic (ES) phases. In the last few years, several deep learning segmentation methods (in particular CNNs) have had great success at estimating these clinical parameters~\cite{Duan19,Bernard2018DeepLT,Oktay17,Zotti18}.  Some of them provide excellent segmentation results with overall Dice index and/or Hausdorff distance within the inter- and intra-observer variations~\cite{Bernard2018DeepLT}. Unfortunately, these methods still generate anatomically impossible shapes like a LV connected to the background or two disconnected RV regions. Therefore, despite their excellent results on average, these methods are still unfit for day-to-day clinical use.

To reduce such errors, several papers integrate shape priors into their cardiac deep learning segmentation methods. In particular, Oktay \mbox{\emph{et al.}} used an approach named anatomically constrained neural network (ACNN)~\cite{Oktay17}. Their neural network is similar to a 3D U-Net, whose segmentation output is constrained to be close to a non-linear compact representation of the underlying anatomy derived from an auto-encoder network. More recently, Zotti \mbox{\emph{et al.}} proposed a method based on the grid-net architecture that embeds a cardiac shape prior to segment MR images~\cite{Zotti18}. Their shape prior encodes the probability of a 3D location point being a member of a certain class and is automatically registered with the last feature maps of their network. Finally, Duan \mbox{\emph{et al.}} implemented a shape-constrained bi-ventricular segmentation strategy \cite{Duan19}. Their pipeline starts with a multi-task deep learning approach that aims to locate specific landmarks. These landmarks are then used to initialize atlas propagation during a refinement stage of segmentation. Although the use of an atlas improves  the quality of the results, their final segmented shapes strongly depend on the accuracy of the located landmarks. From these studies, it appears that only soft constraints are currently imposed in the literature to steer the segmentation outputs towards a reference shape.  As we will be shown in this paper, shape-prior methods are not immune to producing anatomically incorrect results. 

\begin{figure*}[tp]
\centering
\includegraphics[width=0.75\textwidth]{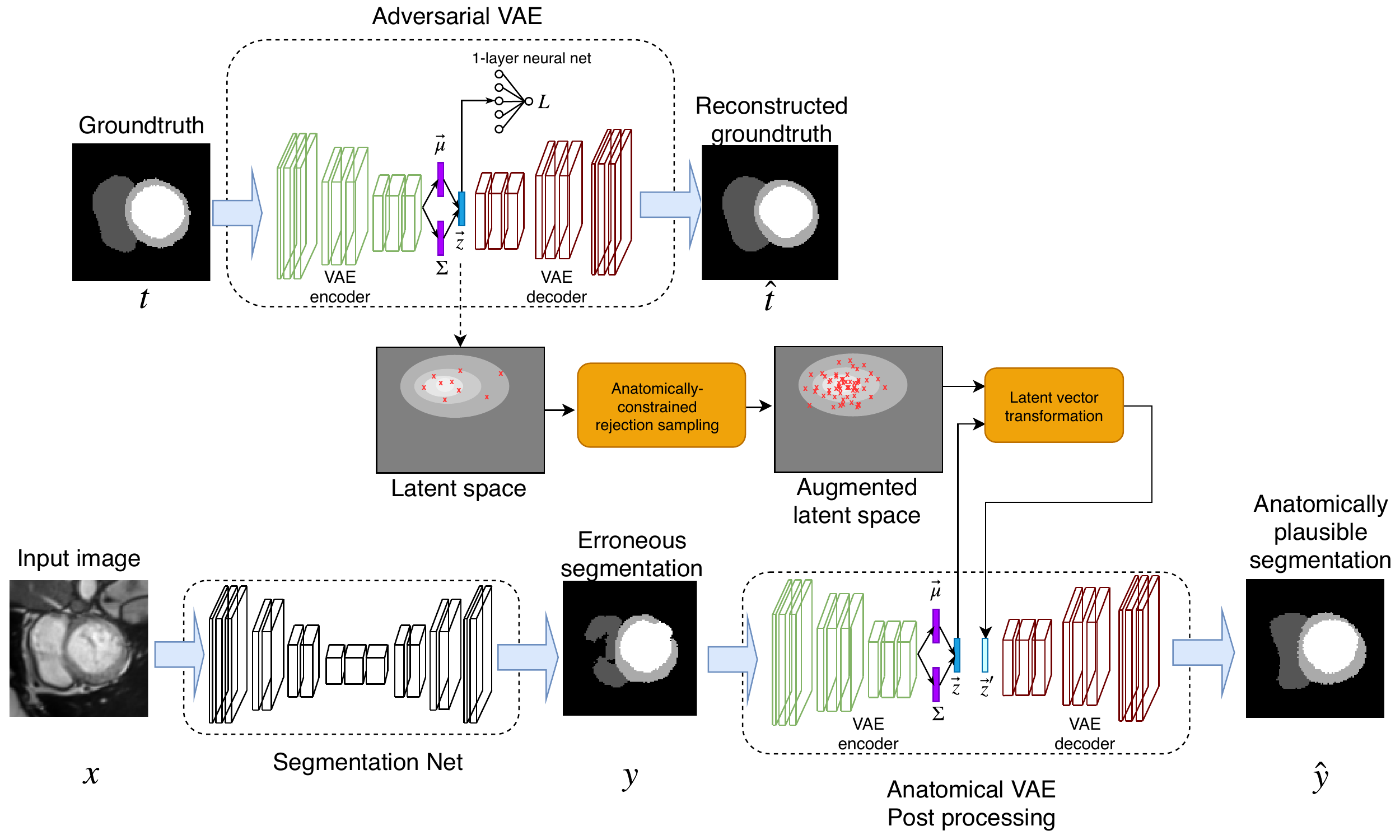}
\caption{Schematic representation of our method.}
\label{fig:ourmethod}
\end{figure*}

Another simple way of reducing the number of anatomically inaccurate results is through the use of post-processing tools.  It typically involves morphological operators or some connected component analysis to remove small isolated regions. Unfortunately, such post-processing methods cannot guarantee the anatomical plausibility of every segmentation map.

In this paper, we present the first deep learning formalism which guarantees the anatomical plausibility of cardiac shapes. Our method can be plugged to the output of any segmentation method as it would reduce to zero its number of anatomically invalid shapes while preserving the overall quality of its results.

\section{Proposed Framework}
As shown in Fig.~\ref{fig:ourmethod}, our method has three main blocks namely: i) an adversarial VAE that learns a  32-dim latent representation of anatomically correct cardiac shapes, ii) an anatomically-constrained data augmentation of the latent vectors and iii) a post-processing VAE which converts erroneous segmentation maps into  anatomically plausible ones.  The anatomical guarantees that our method provides comes from a transformation function that replaces the latent vector of an anatomically erroneous shape by a close but anatomically correct one.

\subsection{Cardiac MR Images and Anatomical Metrics}
\label{sec:metrics}
The goal of our method is to produce cardiac segmentation maps with strong anatomical guarantees from short-axis cine-MRI. In that perspective we defined 16 anatomical metrics that will be used to detect incorrect cardiac shapes.  

We first consider any holes in the LV, the RV or the MYO, and between the LV and the MYO and between the RV and the MYO as being anatomically impossible.  The presence of more than one LV, RV or MYO is also considered implausible.  We also measure if the RV is disconnected from the MYO, if the LV touches the RV or the background, and if the LV, RV and MYO suffer from unusually acute concavities.  The threshold beyond which a concavity is considered abnormal was defined based on the groundtruth of the ACDC training set (c.f. section~\ref{sec::setup}).   We also implemented a circularity metric for the LV and the MYO which is the ratio of their area to that of a circle having the same perimeter.  Again, the threshold for that ratio was obtained from the ACDC training set.  Please note that since these metrics are not included in the loss, they do not need to be differentiable.

\subsection{Adversarial Variational Autoencoder (aVAE)}
VAEs~\cite{kingma2013auto} are encoder/decoder unsupervised learning methods used to derive a latent representation of a set of data.   In our case, the encoder takes as input a cardiac segmentation map $\vec x \in \mathbb{R}^{n\times n}$ and outputs the parameters ($\vec \mu$ and $\vec \sigma$) of a Gaussian probability density $q_\theta(\vec z | \vec x)$ where $\vec z \in \mathbb{R}^{32}$ is a latent vector.  The decoder takes in a latent variable $\vec z$ sampled from $q_\theta(\vec z | \vec x)$ and outputs $\hat{x}$, a reconstructed version of the input cardiac shape $\vec x$.

In our method, we implemented an adversarial VAE (aVAE)~\cite{Makhzani2015AdversarialA} which forces the latent space to be as linear as possible. The constraint comes in the form of a single-layer neural network~\cite{Bishop07} trained simultaneously with the rest of the VAE.  This neural network is used to predict the slice index of the input image $\vec x$ given its latent vector $\vec z$ using a regression loss.  Since the regression's gradient signal propagates through the {\em encoder}, it forces it to learn a more linear (and thus less convoluted) latent space.

\subsection{Anatomically-Constrained Data Augmentation}
\label{sec:augmentation}
Once the aVAE is trained, every groundtruth short-axis cardiac shape is  projected onto the  32d latent space.  Since the ACDC dataset~\cite{Bernard2018DeepLT} contains a total of $1902$ short axis maps, the latent space gets populated by $1902$ latent vectors $\vec z$.  These latent vectors are "anatomically correct" since the deterministic aVAE decoder can convert them back to anatomically valid cardiac shapes.  Unfortunately, these $1902$ vectors are too few to densely populate the 32d manifold of anatomically correct latent vectors.

To solve that problem, we increase the number of anatomically correct latent vectors with a rejection sampling (RS) method~\cite{Koller09}.  The goal is to produce a new set of latent vectors $Z'$ such that the distribution $P(\vec z')$ of the newly generated samples is close to $P(\vec z)$, the distribution from which derive the original $1902$ points.  RS generates a series of samples iid of $P(\vec z)$ but based on a second and easier to sample pdf $Q(\vec z)$.  Since in our case $P(\vec z)$ is unknown, we estimate it with a Parzen window distribution~\cite{Bishop07}. Meanwhile, $Q(\vec z)$ is a Gaussian of mean and variance equal to the distribution of the original $1902$ points.  A key idea with RS is that $P(\vec z) > MQ(\vec z)$ where $M > 1$.  Given $p(\vec z)$ and $Q(\vec z)$, the sampling procedure first generates a random sample $\vec z_i$ iid of $Q(\vec z)$ as well as a uniform random value $u\in [0,1]$.  If $u <  \frac{P(\vec z_i)}{M Q(\vec z_i)}$ then $\vec z_i$ is kept, otherwise it is rejected.  
 
In our case, in addition to an increased number of latent vectors, we want those new vectors to correspond to anatomically correct cardiac shapes.  As such, we redefine the RS criterion as follows:
\begin{eqnarray}
    u < \mathbbm{1}\left ( \mbox{dec}(\vec z_i) \right) \frac{P(\vec z_i)}{M Q(\vec z_i)}
\end{eqnarray}
where $\mathbbm{1}\left ( \mbox{dec}(\vec z_i) \right)$ is an indicator function which returns 1 when the decoded latent vector $\vec z_i$ is a valid cardiac anatomy and zero otherwise.  We call this operation an {\em anatomically-constrained rejection sampling augmentation}.  The procedure is repeated up until the desired number of samples are generated.  This operation allows us to generate 4 million latent vectors which all have a valid cardiac shape, i.e. that respect all 16 metrics defined in Section~\ref{sec:metrics}.  Images of generated samples are provided in the supplementary materials.

\subsection{Latent Vector Transformation}
\label{sec:transformation}
Our system contains a post-processing VAE (at the bottom right of Fig.~\ref{fig:ourmethod}) used to convert erroneous segmentation maps into anatomically valid segmentations.  The post-processing VAE has the same architecture and the same weights as the aVAE.  Thus, any erroneous segmentation map fed to the VAE $encoder$ gets projected into the same latent space as that of the aVAE.  Furthermore, since the VAE $decoder$ is deterministic, any anatomically valid latent vector $\vec z$ is guaranteed to be converted into an anatomically correct cardiac shape.  

The goal is to transform the latent vector $\vec z$ of an erroneous cardiac shape to a similar but anatomically valid latent vector $\vec z'$, which can be summarized as:
\begin{eqnarray}
\vec z'= \arg \min_{\vec z'} ||\vec z - \vec z'||^2, \,\,\,\, s.t.\,\, \mathbbm{1}\left ( \mbox{dec}(\vec z') \right)=1.
\end{eqnarray}
Said otherwise, the goal is to find the anatomically valid latent vector $\vec z'$ that is the closest to $\vec z$.  Unfortunately, since  $\mathbbm{1}\left ( \mbox{dec}(\vec z') \right)=1$ involves the 16 non-differentiable metrics, this function cannot be minimized with a usual Lagrangian formulation.  As a solution, we redefined the problem of finding $\vec z'$ as a problem of finding the smallest vector $\vec \delta_{z'}$ such that $\vec z'=\vec z + \alpha \vec \delta_{z'}$.  In this paper, we recover $\vec \delta_{z'}$ based on the nearest neighbor in the augmented latent space.  In this way,  $\vec \delta_{z'} =(\vec z_{N1}-  \vec z)$ where $\vec z_{N1}$ is the nearest neighbor of $\vec z$ in the augmented latent space and  $\alpha\in [0,1]$.  This leads to an easier 1D optimization problem:
\begin{eqnarray}
\label{eq:optim}
\alpha = \arg \min_{\alpha} |\alpha|, \,\,\,\, s.t. \,\,\mathbbm{1}\left ( \vec z + \alpha \vec \delta_{z'}) \right)=1
\end{eqnarray}
that we solve with a dichotomic search.  At each iteration, the remaining search space of $\alpha$ is divided in two and the anatomical criterion $\mathbbm{1}\left ( \mbox{dec}(\vec z + \alpha \vec \delta_{z'})) \right)=1$ specifies which of the upper-half or lower-half should be divided at the next iteration.  Since the search space decreases exponentially fast, the optimization algorithm is stopped after five iterations,  selecting the smallest alpha that validates the anatomical criterion.

\begin{table}[tp]
    \centering
	\caption{Ablation study of our aVAE showing the average \% of anatomical errors while navigating through the latent space.}
	\begin{tabular*}{0.99\textwidth}{@{\extracolsep{\fill} } ccccc} \hline
	AE & VAE & VAE + registered & VAE + adversarial & VAE + regist. + adv.\\  
	64.76 & 5.84 & 5.85 & 8.48 & 1.25 \\ \hline
	\end{tabular*}
\label{tab:ablationStudyVAE}
\end{table}

\begin{figure}[h]
\centering
\includegraphics[width=0.85\textwidth]{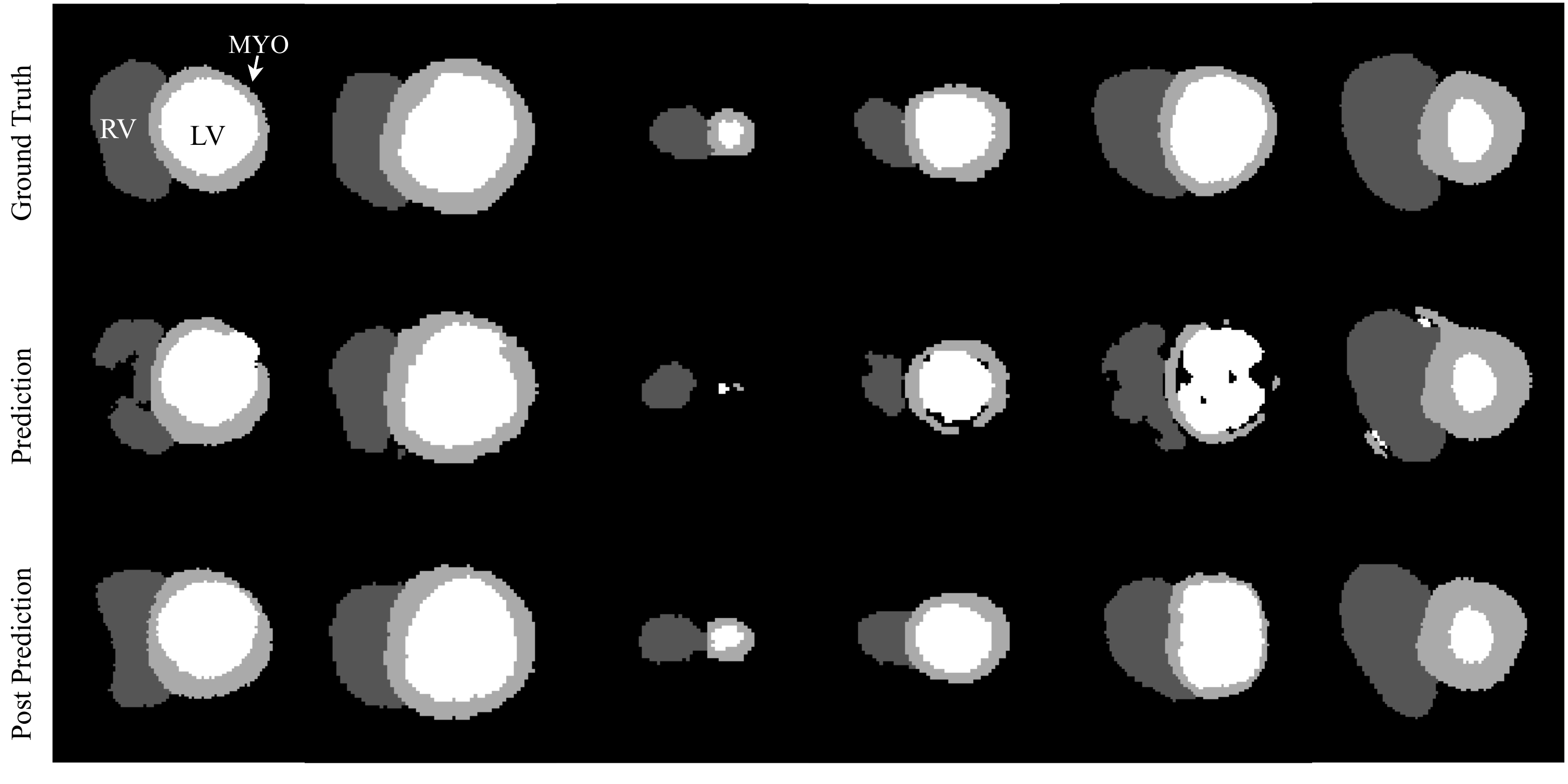}
\caption{\small Groundtruth and erroneous maps before and after our post-processing method.}
\label{fig:result}
\end{figure}

\subsection{Implementation Details}
The {\em encoder} of our aVAE has ten $3\times 3$ convolution layers with stride 2 with ELU~\cite{Clevert2015FastAA} activation layers which output a 32-dim latent vector. The {\em decoder} follows the same architecture except for the transposed convolutions that increase the feature maps resolution.  
For the adversarial network, we used a single-layer neural network with an L2 regression loss.
The whole network is trained end-to-end using Adam~\cite{kingma2014adam}  with a learning rate of $6*10^{-5}$ and a $L2$ weight regularization with $\lambda = 0.01$.  Note that the segmentation maps fed to our VAEs have a size of $256\times 256$ and are registered so the center of the LV is in the middle of the image.  This translation and rotation registration is done at runtime.

\begin{table}[tp]
    \centering
	\caption{Number of anatomically invalid segmentation results on the ACDC test set for 11 segmentation methods with and without our post-processing methods.}
	\smallskip
	\begin{tabular*}{\textwidth}
        {@{} @{\extracolsep{\fill}} l c c ccc @{}}
		\toprule
		\multirow{2}[3]{*}{Submissions} & \multicolumn{1}{c}{Original}
        & \multicolumn{1}{c}{VAE} & \multicolumn{3}{c}{Nearest Neighbors} \\
		\cmidrule(lr){2-2} \cmidrule(lr){3-3} \cmidrule(lr){4-6}
		&&& \mcc{w/o RS} & \mcc{w/ RS} & \mcc{Dicho} \\
		\midrule
        Zotti-2      & 55  & 16  & 0 & 0 & 0 \\
        Khened       & 55  & 16  & 0 & 0 & 0 \\
        Baumgartner  & 79  & 17  & 0 & 0 & 0 \\
		Zotti        & 82  & 15  & 0 & 0 & 0 \\
        Grinias      & 89  & 12  & 0 & 0 & 0 \\
        Isensee      & 128 & 21  & 0 & 0 & 0 \\
        Rohé         & 287 & 40  & 0 & 0 & 0 \\
        Wolterink    & 324 & 42  & 0 & 0 & 0 \\
        Jain         & 185 & 28  & 0 & 0 & 0 \\
        Yang         & 572 & 182 & 0 & 0 & 0 \\ \hline
		ACNN         & 139 & 41  & 0 & 0 & 0 \\ 
		\bottomrule
	\end{tabular*}
\label{tab:ablationStudyAnatomicalMetrics}
\end{table}

\section{Experimental Setup and Results}
\label{sec::setup}

\subsection{Dataset, evaluation criteria, and other methods}
\label{sec::data}
We trained and tested our method on the 2017 ACDC dataset~\cite{Bernard2018DeepLT} which contains cine-MR images of 150 patients, 100 for training and 50 for testing.  As shown in Fig.~\ref{fig:result}, the LV, RV and MYO of every patient has been manually segmented.  We report the average 3D Dice index and Hausdorff distance (HD) for the LV, RV and MYO as well as the LV and RV ejection fraction (EF) absolute error.  Since our approach can accommodate any segmentation method, we tested it on the test results reported by the ten ACDC challengers.  Their methods are summarized by Bernard \mbox{\emph{et al.}}~\cite{Bernard2018DeepLT} except for  Zotti-2~\cite{Zotti18} whose results have been uploaded recently.  We also report results for the ACNN method of Oktay \mbox{\emph{et al.}}~\cite{Oktay17} that uses a latent anatomical prior to train a segmentation CNN. Results from our best implementation (which involves a U-Net and our VAE) are very close to that of the original paper despite the fact that the ACDC training set is smaller than the one they used.  HD values are also slightly larger since we use a 3D HD instead of a 2D HD as in the original paper.

\subsection{Experimental Results}
\label{sec:results}

\subsubsection{Adversarial variational autoencoder}
We validated the design of our aVAE through the ablation study of Table~\ref{tab:ablationStudyVAE}.  Since our post-processing method relies on latent vector interpolation (c.f. Eq~(\ref{eq:optim})), we computed the percentage of anatomically implausible results obtained after interpolating two valid latent vectors.  To do so, we iteratively selected the groundtruth of two random slices from two random patients of the ACDC test set, encoded it to the latent space with the aVAE encoder and linearly interpolated 25 new vectors. We then converted these 25 vectors to segmentation maps with the aVAE decoder and computed their percentage of anatomical errors.  We repeated that process 500 times for the aVAE with and without registration and with and without an adversarial regression loss.  As can be seen, the use of registration and an adversarial regression loss reduces the percentage of anatomically implausible results down to 1.25\% which is more than 4x lower than for the other configurations.

\begin{table}[htb]
    \centering
    \caption{[Top] Average Dice index and Hausdorff distance (in mm) and [Bottom] Average error on LV and RV ejection fraction (EF) for the ACDC test set with and without our post-processing method.}
    \smallskip
    \begin{tabular*}{\textwidth}
        {@{} @{\extracolsep{\fill}} l c c ccc @{}}
        \toprule
        \multirow{2}[3]{*}{Submissions} & \multicolumn{1}{c}{Original}
        & \multicolumn{1}{c}{VAE} & \multicolumn{3}{c}{Nearest Neighbors} \\
		\cmidrule(lr){2-2} \cmidrule(lr){3-3} \cmidrule(lr){4-6}
		&&& \mcc{w/o RS} & \mcc{w/ RS} & \mcc{Dicho} \\
        \midrule
        Zotti-2      & .913/9.7  & .910/10.1 & .899/14.4 & .909/11.0 & .910/10.1 \\
        Khened       & .915/11.3 & .912/12.3 & .894/15.2 & .909/12.7 & .912/10.9 \\
        Baumgartner  & .914/10.5 & .911/11.2 & .889/18.2 & .907/12.6 & .910/10.6 \\
        Zotti        & .910/9.7  & .907/10.9 & .878/19.6 & .903/12.6 & .907/11.0 \\
        Grinias      & .835/15.9 & .833/19.3 & .752/32.5 & .825/16.9 & .833/15.8 \\
        Isensee      & .926/9.1  & .923/10.7 & .881/18.4 & .917/11.2 & .923/9.2  \\
        Rohé         & .891/12.2 & .887/14.6 & .756/32.2 & .874/15.1 & .887/12.8 \\
        Wolterink    & .907/10.8 & .903/13.0 & .752/32.8 & .887/13.5 & .903/11.0 \\
        Jain         & .891/12.2 & .886/12.6 & .820/31.9 & .878/14.2 & .886/11.6 \\
        Yang         & .800/27.5 & .752/21.7 & .455/29.7 & .722/11.5 & .752/10.2 \\ \hline
        ACNN         & .892/12.3 & .886/26.2 & .885/12.0 & .885/12.2 & .889/13.1\\ 
        \bottomrule 
        \toprule
        Zotti-2      & 2.54/5.11 & 2.63/5.12 & 2.49/5.57 & 2.58/5.18 & 2.62/5.18 \\
        Khened       & 2.39/5.24 & 2.41/4.96 & 2.70/5.36 & 2.63/5.07 & 2.42/5.27 \\
        Baumgartner  & 2.58/6.00 & 2.62/6.30 & 2.83/6.72 & 2.85/6.48 & 2.64/6.33 \\
        Zotti        & 2.98/5.48 & 2.98/5.42 & 3.06/5.72 & 3.10/5.71 & 3.06/5.59 \\
        Grinias      & 4.14/7.39 & 4.18/7.86 & 4.67/8.00 & 4.33/7.35 & 4.01/7.43 \\
        Isensee      & 2.16/4.85 & 2.15/4.61 & 2.49/5.58 & 2.35/4.48 & 2.20/4.82 \\
        Rohé         & 2.84/8.18 & 2.95/7.85 & 3.13/8.93 & 3.39/7.97 & 2.91/8.11 \\
        Wolterink    & 2.75/6.59 & 2.82/6.39 & 3.40/6.93 & 3.48/6.07 & 2.84/6.44 \\
        Jain         & 4.36/8.49 & 4.35/8.83 & 4.98/9.63 & 4.59/8.69 & 4.40/8.72 \\
        Yang         & 6.22/15.99 & 6.80/20.56 & 7.57/27.9 & 7.77/22.09 & 9.10/21.76 \\ \hline
        ACNN         & 2.46/3.68 & 2.53/4.09 & 2.51/3.89 & 2.96/3.82 & 2.50/3.71 \\ 
        \bottomrule
    \end{tabular*}
\label{tab:ablationStudySimilarityMetrics}
\end{table}

\subsubsection{Postprocessing results}
Results on the ACDC test set are in Table~\ref{tab:ablationStudyAnatomicalMetrics}, Table~\ref{tab:ablationStudySimilarityMetrics}. Table~\ref{tab:ablationStudyAnatomicalMetrics} contains the total number of slices with at least one anatomical error, Table~\ref{tab:ablationStudySimilarityMetrics} shows at the top the overall Dice index and HD and at the bottom the LV and RV EF absolute errors.  Results without our post-processing are under the {\em Original} column.  As shown, every method produces a non-negligible number of anatomical errors (the ACDC testset has a total of $1078$ slices).

By feeding every erroneous segmentation map to our VAE without transforming the latent vector $\vec z$, we get to drastically reduce the number of anatomical errors without affecting too much the HD, the Dice and the EF.  This comes as no surprise since the VAE was trained to reproduce groundtruth (and thus anatomically correct) cardiac shapes.  However, like any neural network, a basic VAE provides no guarantee on the quality of its output.  To completely eliminate erroneous segmentations, we first swap erroneous latent vectors with their nearest neighbor (i.e. by fixing $\alpha$ to 1 in Eq.~\ref{eq:optim}) using groundtruth data from the ACDC training set (i.e. $1902$ short axis maps) without RS augmentation (w/o RS).  While that procedure eliminated every anatomical error and did not change much the EF error, the Dice index and HD suffered considerably.  We then tested the same method but with the latent space augmented by 4 million anatomically correct vectors (c.f. Section~\ref{sec:augmentation}).  This approach (w/ RS) also provides strong anatomical guarantees but better Dice and HD than without RS.  The last column shows the results of our complete method, i.e.  Eq.(~\ref{eq:optim}) optimized with a dichotomic search.   While results are all anatomically correct, the EF error, the Dice index are almost identical to that of the original methods.  The HD also never increases more than 1.3mm, which, considering that the average voxel size is near 1.4x1.4x10mm3, corresponds to less than 1 pixel in the image.  This shows that our approach does not degrade the overall results but only warps anatomically incorrect results towards the closest anatomically viable shape.  Fig.~\ref{fig:result} shows erroneous predictions before and after our post-processing.  While the correct areas are barely affected by our method, erroneous sections, big or small, get smoothly warped.  Our method takes roughly 1 sec to process a 2D image on a mid-end computer equipped with a Titan X GPU.

\section{Conclusion}
\label{sec::conclu}
We presented a post-processing VAE which converts anatomically invalid cardiac shapes into close but correct shapes.  Our method relies on 16 anatomical metrics that we use both to detect abnormalities and populate an aVAE latent space.  Since those metrics are not included in the loss, they need not be differentiable.  According to the inter- and intra-expert variations reported by Bernard \mbox{\emph{et al.}}\cite{Bernard2018DeepLT}, methods such as Isensee \mbox{\emph{et al.}}, Zotti-2, Khened and Baumgartner are on average as accurate as an expert and, with our post-processing method, are now guaranteed to produce anatomically plausible results.
This method will be integrated in the cardiac images processing software QIR (\url{https://www.casis.fr/}).

\section{Acknowledgments}
\label{sec::ack}
We would like to extend our thanks to CASIS - CArdiac Simulation \& Imaging Software (\url{https://www.casis.fr/}) company for their support.

%
%
{

\bibliographystyle{plain}
\footnotesize
\bibliography{paper1513}

}
\end{document}